\documentclass[oldversion]{aa}
\usepackage{graphicx}
\usepackage{color}
\usepackage{txfonts}
\usepackage{longtable}

\setlongtables
\begin{document}
\title{Multi-epoch VLBA observations of 3C 66A }
\author{  H.-B. Cai    \inst{1,2}
       \and Z. -Q. Shen     \inst{1,3}
       \and H. Sudou \inst{4}
       \and L.-L. Shang \inst{1}
       \and S. Iguchi \inst{5}
       \and Y. Murata \inst{6}
       \and Y. Taniguchi \inst{7}
       \and K. Wakamatsu \inst{4}
       \and H. Takaba \inst{4}
       }%
\offprints{Z.-Q. Shen, \email{zshen@shao.ac.cn}} \institute{Shanghai Astronomical Observatory, Chinese Academy
of Sciences, Shanghai 200030, China \and Graduate School of Chinese Academy of Sciences, Beijing 100039, China
\and Joint Institute for Galaxy and Cosmology, SHAO and USTC, China \and Faculty of Engineering, Gifu
University, Gifu 501-1193, Japan \and National Astronomical Observatory of Japan, 2-21-1 Osawa, Mitaka, Tokyo
181-8588, Japan \and The Institute of Space and Astronautical Science, Japan Aerospace Exploration Agency, 3-1-1
Yoshinodai, Sagamihara, Kanagawa 229-8510, Japan \and Physics Department, Graduate School of Science, Ehime
University, 2-5 Bunkyou, Matuyama, 790-8577, Japan}

\date{Received 30 January 2007 / Accepted 22 March 2007}

\abstract {We present the results of six-epoch Very Long Baseline Array (VLBA) observations of 3C~66A. The
high-resolution Very Long Baseline Interferometer (VLBI) maps obtained at multi-frequency (2.3, 8.4, and 22.2
GHz) simultaneously enabled us to identify the brightest compact component with the core. We find that the
spectrum of the core can be reasonably fitted by the synchrotron self-absorption model. Our VLBA maps show that
the jet of 3C~66A has two bendings at about 1.2 and 4 mas from the core. We also give possible identifications
of our jet components with the components in previous VLBA observations by analysing their proper motions. We
find consistent differences of the position from the core in one component between different frequencies at six
epochs.

}
\maketitle

\section{Introduction}

The source 3C 66A (B0219+428, 4C 42.07) is a low-frequency peaked BL Lac object (LBL) peaking at IR-UV wavebands
(e.g., Perri et al. \cite{Perri03}; Joshi \& B\"{o}ttcher \cite{Joshi06}). Since its optical counterpart was
identified by Wills \& Wills (\cite{Wills74}), a large number of observations from radio to gamma-ray bands have
been performed on 3C~66A. As a blazar, 3C~66A also exhibits a prominent variability at radio, IR, and optical
(cf. B\"{o}ttcher et al. \cite{Bottcher05}, hereafter B05).

Price et al. (\cite{Price93}) gave the first Very Long Array (VLA) maps at 1.5 and 5 GHz, both showing that 3C
66A has a $6^{''}$ extended structure along the position angle (P.A.) $\sim$ 170$\degr$. New VLA maps at 1.5 and
5 GHz (Taylor et al. \cite{Taylor96}) further revealed that 3C 66A has two weak lobes, one at $15^{''}$ south of
the central region, the other at $10^{''}$ with P.A. about $-20\degr$. Taylor et al. (\cite{Taylor96}) present
the first 5 GHz VLBA map, exhibiting a typical core-jet structure in 3C~66A. Jorstad et al. (\cite{Jorstad01},
hereafter J01) studied the jet's kinematics of 3C 66A through multi-epoch VLBA observations at 22 and 43 GHz,
and they detected superluminal motions for four jet components with the apparent velocities from 15.3c to 29.7c
(the apparent velocities shown in this paper have been calculated by using the cosmological parameters given
below), and one stationary jet component. A more detailed kinematic study of 3C~66A was done by Jorstad et al.
(\cite{Jorstad05}, hereafter J05) through the total and polarized intensity VLBA observations at 43 GHz at 17
epochs spanning 1998 March to 2001 April, with some epochs accompanied by nearly simultaneous polarization
measurements at 100, 222/353 GHz, and the optical wavebands. B05 presented a coordinated observation of 3C~66A
during the Whole Earth Blazar Telescope (WEBT) campaign at multiple frequencies including radio, infrared,
optical, X-rays, and gamma-rays at very high energy. Their VLBA observations at three epochs at 22 and 43 GHz
revealed superluminal motion for only one jet component with an apparent velocity of 12.1c, other jet components
remain stationary, and the radial radio brightness profile suggests a magnetic field decay $\propto$ r$^{-1}$
and, thus, a predominantly perpendicular magnetic field orientation.

We describe in Sect. 2 our VLBA observations of 3C~66A at relatively low frequencies of 2.3 and 8.4 GHz during
six epochs and at 22.2 GHz during the first two of six epochs. The results are presented in Sect. 3. We analyse
the spectra of the VLBI components in Sect. 3.1, study the kinematics of the jet components in Sect. 3.2, and
the positional difference of the jet components caused by the frequency-dependent core position offset in Sect.
3.3. The discussion of results is given in Sect. 4, followed by conclusions in Sect. 5.

We adopt 0.444 as the redshift of 3C~66A (Miller et al. \cite{Miller78}), although Bramel et al.
(\cite{Bramel05}) thought that this redshift value is still uncertain and deserves more spectroscopic
observations. Throughout this paper the radio spectral index $\alpha$ is defined as
$S_{\nu}\propto\nu^{\alpha}$. By assuming $H_0=71$~km~s~$^{-1}$~Mpc$^{-1}$, $\Omega_M=0.27$, and
$\Omega_\Lambda=0.73$ (Spergel et al. \cite{Spergel03}), we have a scale for 3C~66A (z = 0.444) of 1 mas $=5.69
$ pc.

\section{Observations and data reduction}

The observations of 3C~66A were performed with the NRAO\footnote{The National Radio Astronomy Observatory (NRAO)
is operated by Associated Universities Inc., under cooperative agreement with the National Science Foundation.}
VLBA at 2.3 and 8.4 GHz during the following six epochs: 2001 March 13-14 (2001.20), 2001 June 25 (2001.48),
2001 November 9 (2001.86), 2002 February 8-9 (2002.11), 2002 February 21-22 (2002.14), and 2002 June 14
(2002.45). The VLBA observations at 22.2 GHz were successfully performed at the above first two epochs. These
observations were used to search for the binary black hole in 3C~66B (Sudou et al. \cite{Sudou03}). The
on-source time at each epoch is about 30 min at 2.3 and 8.4 GHz, and 70 min at 22.2~GHz. The recording bandwidth
is 16 MHz (two 8 MHz IF channels) at 2.3 and 8.4 GHz, and 32 MHz (four 8 MHz IF channels) at 22.2 GHz. All the
data were recorded with the right-circular polarization (RCP) mode at 2.3 and 8.4 GHz and with the left-circular
polarization (LCP) mode at 22.2 GHz.

\begin{figure*}
\scalebox{0.6}[0.6]{\rotatebox{0}{\includegraphics{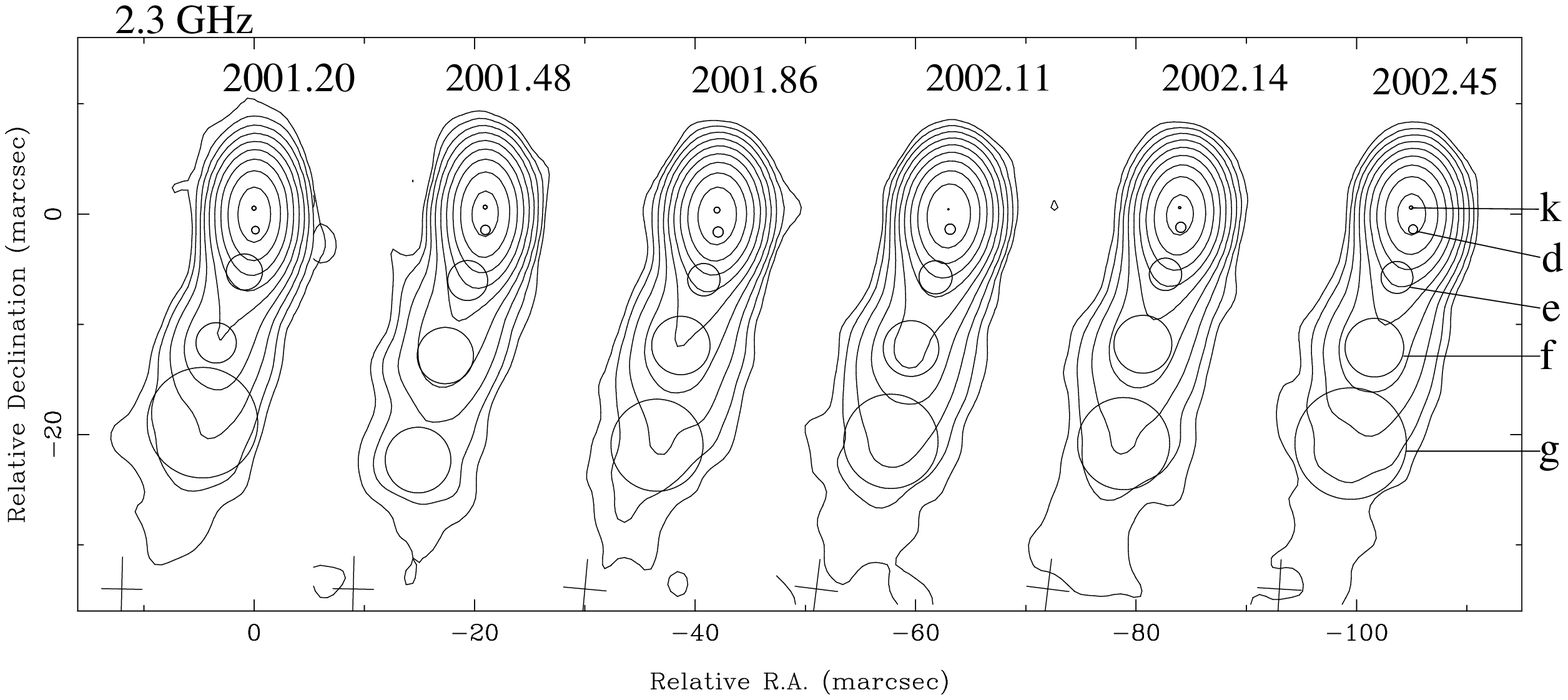}}}
\scalebox{0.6}[0.6]{\rotatebox{0}{\includegraphics{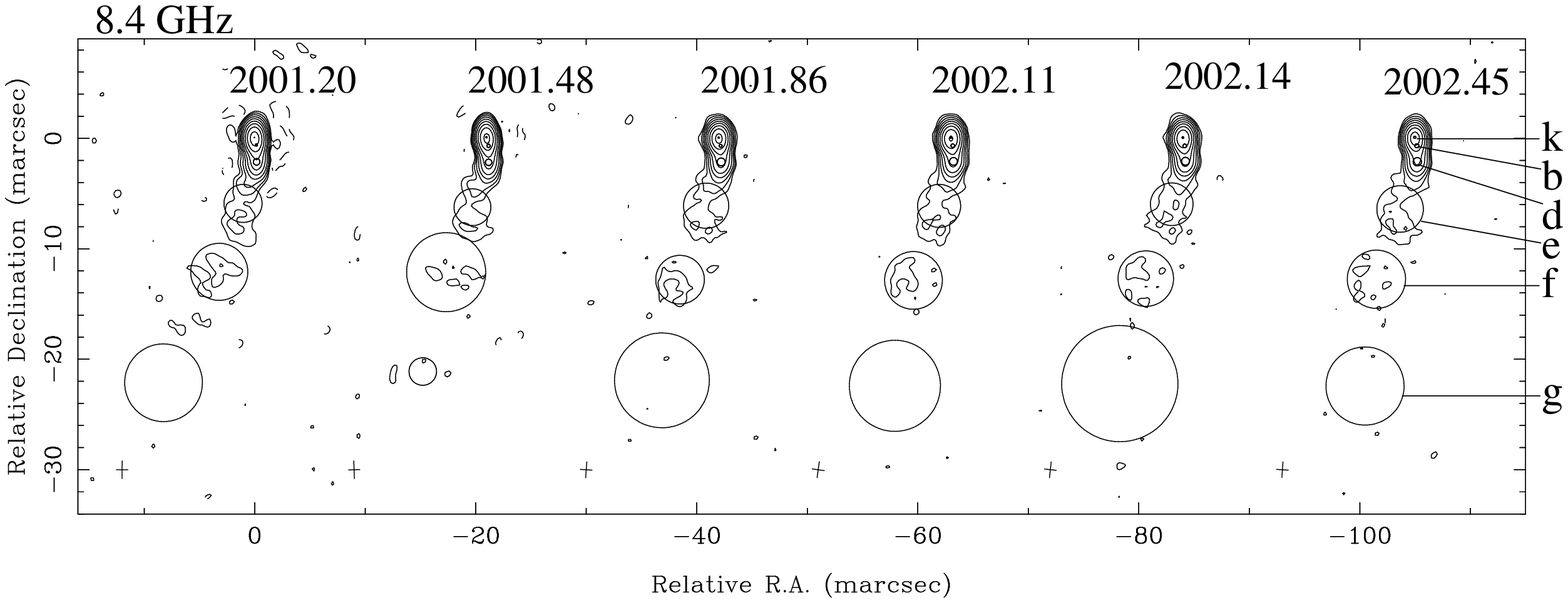}}}
\scalebox{0.6}[0.6]{\rotatebox{0}{\includegraphics{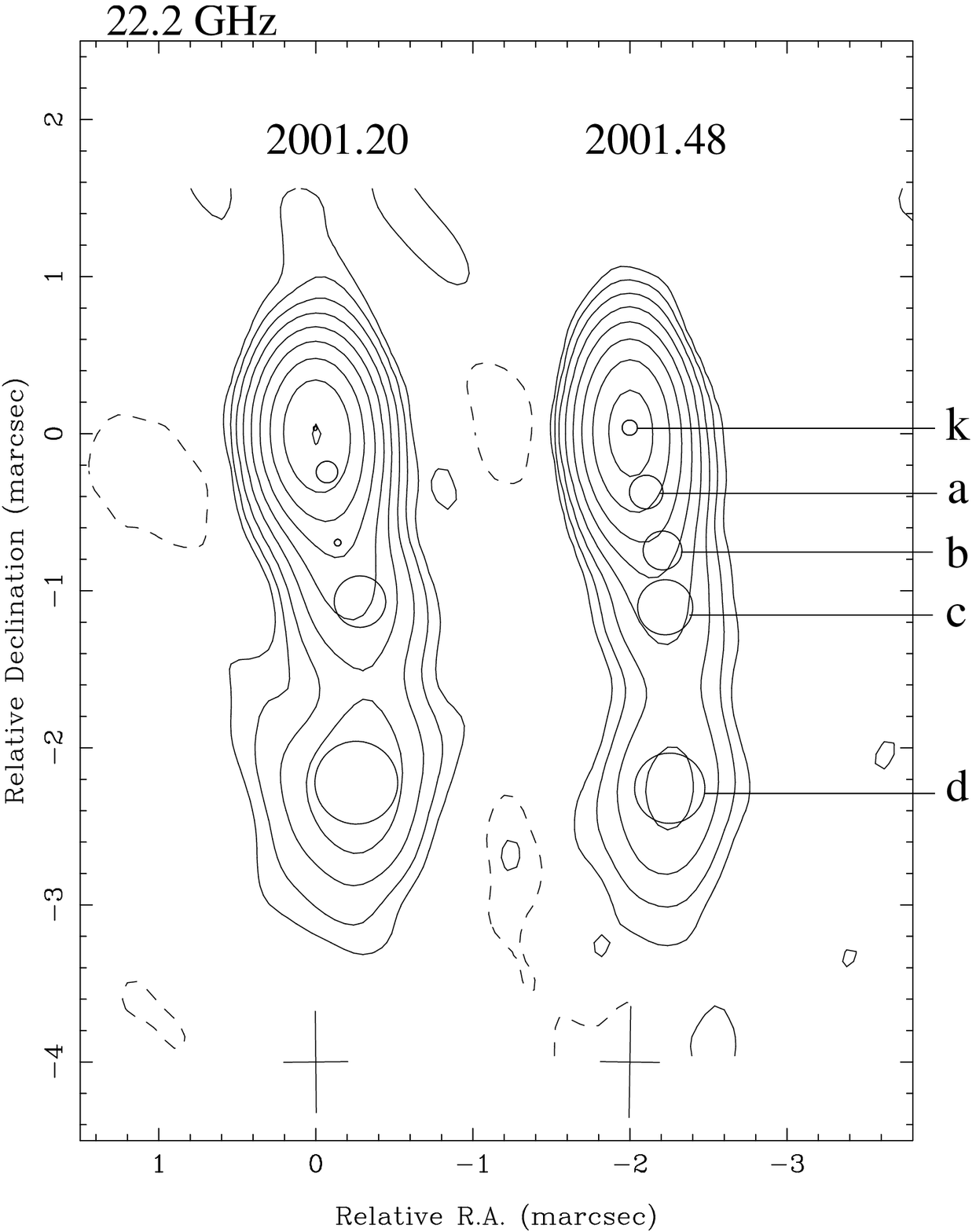}}} \caption{The naturally-weighted VLBA maps of
3C~66A at 2.3, 8.4, and 22.2 GHz from top to bottom. The circles superimposed on the maps represent the Gaussian
model components listed in Table 2, whose names are labelled. The crosses represent the restoring beams in Table
1.} \label{fig:1}
\end{figure*}

The data correlation was done with the VLBA correlator in Socorro, New Mexico, USA. Post-correlation data
analysis was done in the NRAO AIPS software (Schwab \& Cotton \cite{Schwab83}) for fringe fitting and in the
Caltech DIFMAP package (Shepherd \cite{Shepherd97}) for hybrid mapping. A prior visibility amplitude calibration
was done using the antenna gain and the system temperature measured at each station. During the process of
amplitude calibration, the correction for atmospheric opacity was also applied at all the three frequencies.
During the hybrid mapping, the CLEAN and phase-only self-calibration were iteratively used for the early
processing, and the amplitude self-calibration was added for the later processing. The final VLBA maps are shown
in Fig.~\ref{fig:1}. The mapping parameters of Fig.~\ref{fig:1} are listed in Table~\ref{tab:1}. The
quantitative description of the source structure was determined by the circular Gaussian model fitting to the
calibrated visibility data, and the results are listed in Table~\ref{tab:2}.

\begin{table*}[]
\centering
\begin{minipage}{140mm}
  \caption[]{Description of VLBA maps of 3C~66A shown in Fig.~\ref{fig:1} }
  \label{tab:1}
  \begin{tabular}{cccccccc}
\hline\noalign{\smallskip} \hline\noalign{\smallskip} &&&\multicolumn{3}{c}{Restoring Beam}\\ \cline{4-6}
   $\nu$ & Epoch      & S$_{peak}$      & Major     & Minor     & P.A.& Contours   \\
   (GHz)  & (yr)   & (Jy/beam) &(mas)   &(mas)   &(deg)&(mJy/beam)  \\
   (1)    &(2)        &(3)        &(4)     &(5)     &(6)  &(7)        \\
  \hline\noalign{\smallskip}
2.3 GHz   & 2001.20    & 0.678      & 5.78    & 3.67 &-1.21&1.70$\times$(-1,1,2,4,8,16,32,64,128,256)  \\
   &2001.48     & 0.641      & 5.96    & 3.68 &-1.24&1.88$\times$(-1,1,2,4,8,16,32,64,128,256)  \\
   &2001.86     & 0.748      & 5.39    & 3.87 &-5.16&1.70$\times$(-1,1,2,4,8,16,32,64,128,256)  \\
   &2002.11     & 0.718      & 5.4    & 3.9 &-7.06&1.42$\times$(-1,1,2,4,8,16,32,64,128,256)  \\
   &2002.14     & 0.598      & 5.47    & 3.87 &-7.1&1.79$\times$(-1,1,2,4,8,16,32,64,128,256)  \\
   &2002.45     & 0.584      & 5.33    & 3.98 &-3.55&1.72$\times$(-1,1,2,4,8,16,32,64,128,256) \\\hline
8.4 GHz   & 2001.20    & 0.653      & 1.6    & 1.04 &-0.34&1.74$\times$(-1,1,2,4,8,16,32,64,128,256)  \\
   &2001.48     & 0.719      & 1.61    & 1.01 &-2.06&2.34$\times$(-1,1,2,4,8,16,32,64,128,256)  \\
   &2001.86     & 0.696      & 1.44    & 1.08 &-4.23&2.18$\times$(-1,1,2,4,8,16,32,64,128,256)  \\
   &2002.11     & 0.623      & 1.46    & 1.09 &-7.81&2.27$\times$(-1,1,2,4,8,16,32,64,128,256)  \\
   &2002.14     & 0.586      & 1.43    & 1.07 &-6.05&2.28$\times$(-1,1,2,4,8,16,32,64,128,256)  \\
   &2002.45     & 0.494      & 1.42    & 1.06 &-4.22&2.63$\times$(-1,1,2,4,8,16,32,64,128) \\\hline
22.2 GHz   & 2001.20    & 0.718      & 0.64    & 0.41 &0.57&2.73$\times$(-1,1,2,4,8,16,32,64,128,256)  \\
   &2001.48     & 0.638      & 0.71    & 0.38 &-0.80&3.50$\times$(-1,1,2,4,8,16,32,64,128)  \\
  \noalign{\smallskip}\hline
  \end{tabular}
\\
Notes:(1) Observing frequency;
      (2) Observing epoch;
      (3) Peak flux density;
      (4), (5), (6) Parameters of the restoring Gaussian beam: the full
width at half maximum (FWHM) of the major and minor axes and the position angle (P.A.) of the major axis.
      (7) Contour levels of the map. The lowest contour level is three times the rms noise in the
      maps.
      \end{minipage}
\end{table*}

For the VLBI spectral analysis, an accurate absolute calibration of the flux density is necessary. Unfortunately
due to the lack of the corresponding VLBA observations for other compact amplitude calibrators, we could not
perform any absolute amplitude calibration. Based on our experience of the study on PKS~0528+134 and 3C~138 by
the multi-frequency VLBA observations (Cai et al. \cite{Cai06}; Shen et al. \cite{Shen05}), the errors in the
absolute flux-density calibration are about 10\% at 2.3 and 8.4 GHz, and about 20\% at 22.2 GHz. We also used
the Difwrap program (Lovell \cite{Lovell00}) to estimate the errors of the Gaussian model-fitting parameters.
The errors of the flux densities in the Gaussian model fitting estimated by Difwrap are much smaller than the
errors in the absolute flux-density calibration, so we only adopted the errors in the absolute flux-density
calibration as the error bars of flux density, and adopted the errors of other model-fitting parameters
estimated by Difwrap. All the errors are listed in Table~\ref{tab:2}.

\section{Results}

The fitted VLBI components listed in Table~\ref{tab:2} are also labelled in Fig.~\ref{fig:1}. We label the
component at the north end (having the smallest size and the highest brightness temperature) k component, which
is further identified with the core according to its spectral characteristics in Sect. 3.1. Other jet components
are labelled alphabetically according to their separations from the k component.

Component c detected at 22.2 GHz was not detected at 2.3 and 8.4 GHz, which may be due to the lower resolution
at 2.3 and 8.4 GHz and its relative weakness at 2.3 and 8.4 GHz when compared with components b and d. The maps
at 8.4 and 22.2 GHz show that there are two bendings in the jet emission, at about 1.2 and 4 mas from component
k, with an extended jet structure toward P.A. $\sim$ 170$\degr$ at 20-30 mas. This extension toward the south is
consistent with that of the VLA maps at 1.5 and 5 GHz (Price et al. \cite{Price93}; Taylor et al.
\cite{Taylor96}) and the VLBA map at 5 GHz in Taylor et al. (\cite{Taylor96}). But the structure toward the
north that has been detected in the VLA maps at 1.5 and 5 GHz is not shown in our VLBA maps and the VLBA map at
5 GHz either (Taylor et al. \cite{Taylor96}). This may be due to the lobe in the north (P.A. $\sim$ -20\degr),
detected by VLA at 1.5 and 5 GHz, being too weak and dispersive, and thus resolved by the VLBA. In the following
subsections, we study the spectra, motions, and the positional offsets of these jet components based on our VLBA
observations.

\subsection{Spectra of VLBI components and identification of the core}

\begin{figure*}
   \scalebox{0.8}[0.8]{\includegraphics[0,0][300,235]{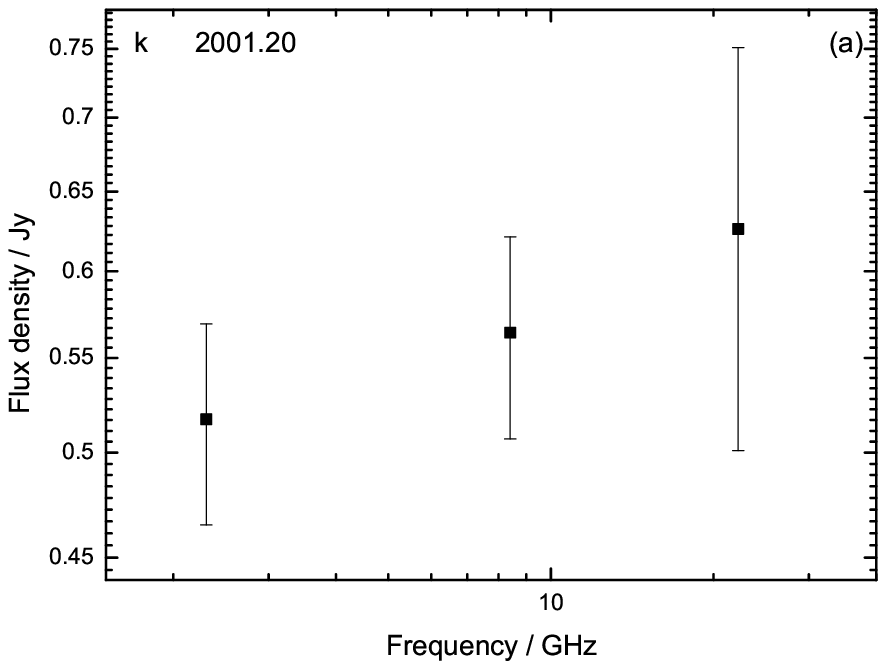}}
   \includegraphics[width=85mm]{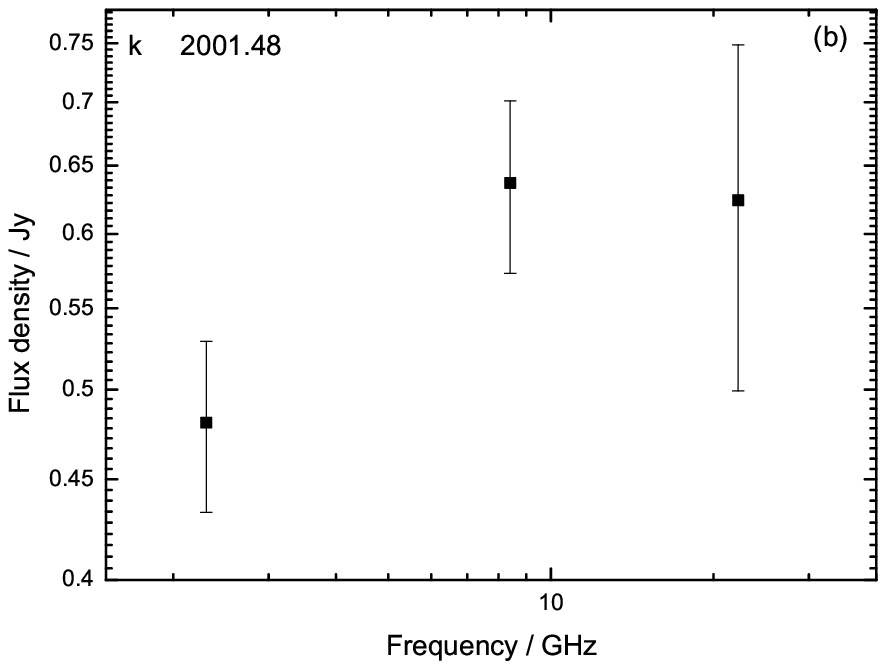}
   \includegraphics[width=85mm]{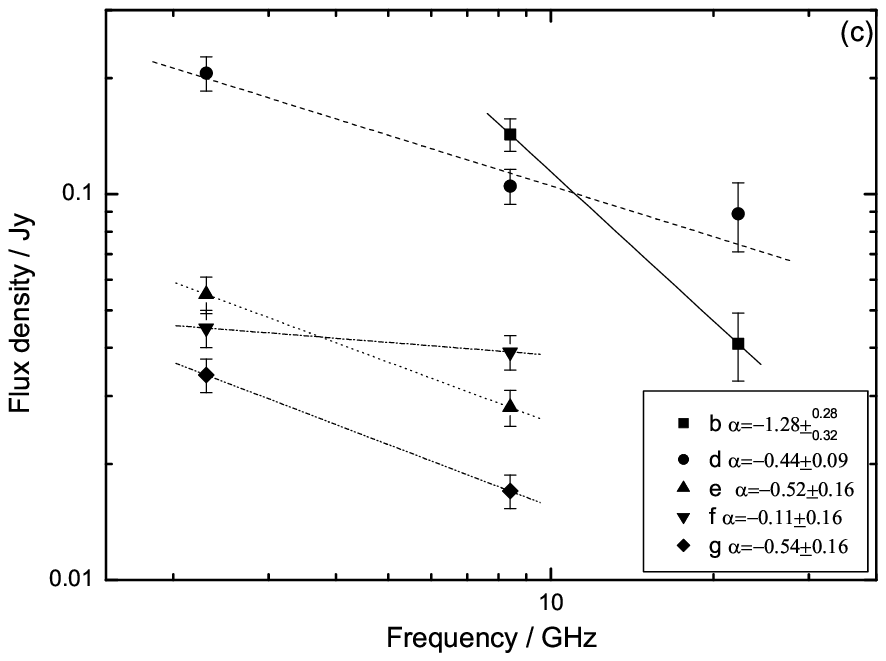}
   \includegraphics[width=85mm]{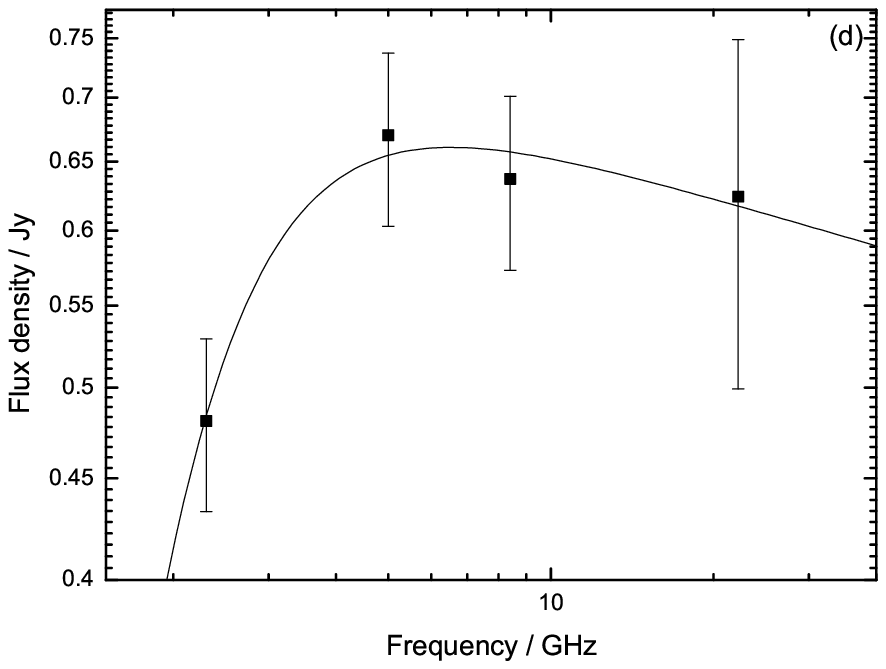}
\caption{The spectra of VLBI components in 3C 66A: (a) and (b) are the spectrum of core (k component) at 2001.20
and 2001.48, and (c) is the spectra of the jet components b, d, e, f, and g at 2001.20. Shown in (d) is an SSA
fitting curve to the spectrum of k component with all the data points from our observations at 2001.48 except 5
GHz data from Taylor et al. (\cite{Taylor96}).}
   \label{fig:2}
   \end{figure*}

The detection of components (k, b, d, e, f, and g) at more than one frequency allows spectral analysis of these
VLBI components. We plot the spectrum of k component in Fig.~\ref{fig:2} (a) and (b), and the other five
components' spectra in Fig.~\ref{fig:2} (c). The data in Fig.~\ref{fig:2} (a), (b), and (c) are from our
observations at 2001.20, 2001.48, and 2001.20, respectively. The spectral indexes of component d at the first
two epochs (2001.20 and 2001.48) were obtained from the linear regression on the data points at 2.3, 8.4, and
22.2 GHz. The spectral indexes of the other components are simply from the two-point measurements at 2.3 and 8.4
GHz, except for component b at the first two epochs whose spectral indexes come from two-point measurements at
8.4 and 22.2 GHz. The results from these spectral fits during six epochs are given in Table~\ref{tab:3}, where
we can see that the k component's spectral indexes, different from other components, are positive, implying the
existence of an absorption. Its inverse spectrum at the low frequency in Fig.~\ref{fig:2} (b) also confirms
this. Combined with its compactness, we identify k as the core of 3C~66A.


\addtocounter{table}{1}
\setlength{\tabcolsep}{3pt}
\begin{table}
\begin{minipage}{90mm}
\begin{center}\caption{The spectral indexes of the VLBI components.} \label{tab:3} \vspace{2mm}
\begin{tabular}{ccccccc}\hline\hline
 &\multicolumn{6}{c}{component}\\ \cline{2-7}
epoch&k& b& d& e& f&g \\\hline 2001.20&
0.067&-1.28$\pm^{0.28}_{0.32}$&-0.44$\pm$0.09&-0.52&-0.11&-0.54\\
2001.48&0.22&-4.52$\pm(>0.32)$&-0.47$\pm$0.09&-0.28&0&-2.27$\pm(>0.47)$\\
2001.86&0.052&&-0.51&-0.11&-0.48&-0.19\\
2002.11&0.05&&-0.57&-0.24&-0.23&-0.54\\
2002.14&0.21&&-0.56&-0.15&-0.32&-0.11\\
2002.45&0.03&&-0.41&-0.12&-0.17&-0.30\\\hline
\end{tabular}
\end{center}
Note: The errors, if not shown in the table, are all 0.16.
\end{minipage}
\end{table}

Taylor et al. (\cite{Taylor96}) report on the VLBI observations of 3C~66A at 5 GHz at 1995.65 with a core flux
density 0.67 Jy. According to the single-dish monitoring by the University of Michigan Radio Observatory
(UMRAO), the total flux densities of 3C 66A at 5 GHz at epochs 2001.20 and 2001.48 of 2.25 and 2.37 Jy,
respectively, are almost the same as that at epoch 1995.65 of 2.38 Jy (Taylor et al. \cite{Taylor96}). By
assuming that the same total flux density infers the same core activity, we can estimate the flux density of
VLBI core to be 0.67 Jy at 5 GHz at 2001.48 (as plotted in Fig. 2 (d)). Then we use the synchrotron-self
absorption (SSA) formula $S_{\nu}=S_0\nu^{2.5}[1-{\rm exp}(-\tau_s\nu^{\alpha-2.5})]$ to fit the data at 2.3, 5,
8.4, and 22.2 GHz, where $S_0$ is the flux density in Jy at 1 GHz when the SSA optical depth at 1 GHz $\tau_s$
$\gg1$ and $\alpha$ is the optically thin spectral index. The best-fit $\alpha=-0.08$, $S_0=0.10$ Jy, and
$\tau_s=7.92$, the fitting curve is plotted in Fig. 2 (d). From the fitting curve we estimate that the peak flux
density $S_m$ is $\sim$0.66 Jy at the inverse frequency $\nu_m=6.52$ GHz. Using the relation $\frac{\nu_m}{1~
{\rm GHz}}\sim8(\frac{B}{1~{\rm G}})^{\frac{1}{5}}(\frac{S_m}{1~{\rm Jy}})^{\frac{2}{5}}(\frac{\theta}{1~{\rm
mas}})^{-\frac{4}{5}}(1+z)^{\frac{1}{5}}$ for the SSA model (Kellermann \& Pauliny \cite{Kellermann81}), we
estimate the magnetic field in the core region $\sim$1.1 mG, here $\theta$, the angular diameter of the core of
0.21 mas is the geometrical mean of the angular diameters at 2.3 and 8.4 GHz. For comparison, the angular size
of 3C~66A at 5 GHz is 0.2 mas from previous VLBA observations (Taylor et al. \cite{Taylor96}).

\subsection{Proper motion and registration of jet components}

\begin{figure*}
   \includegraphics[width=85mm]{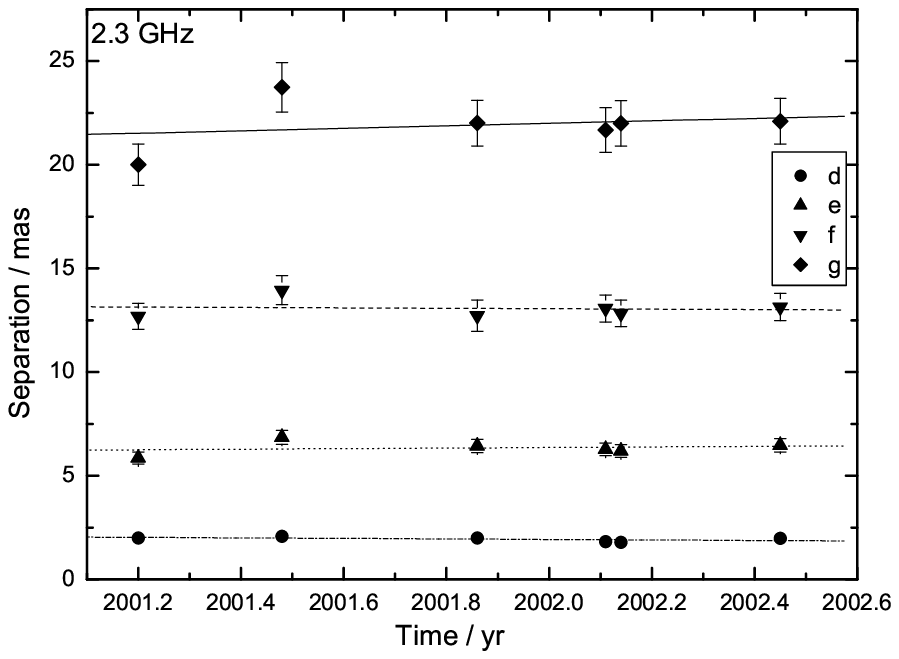}
   \includegraphics[width=85mm]{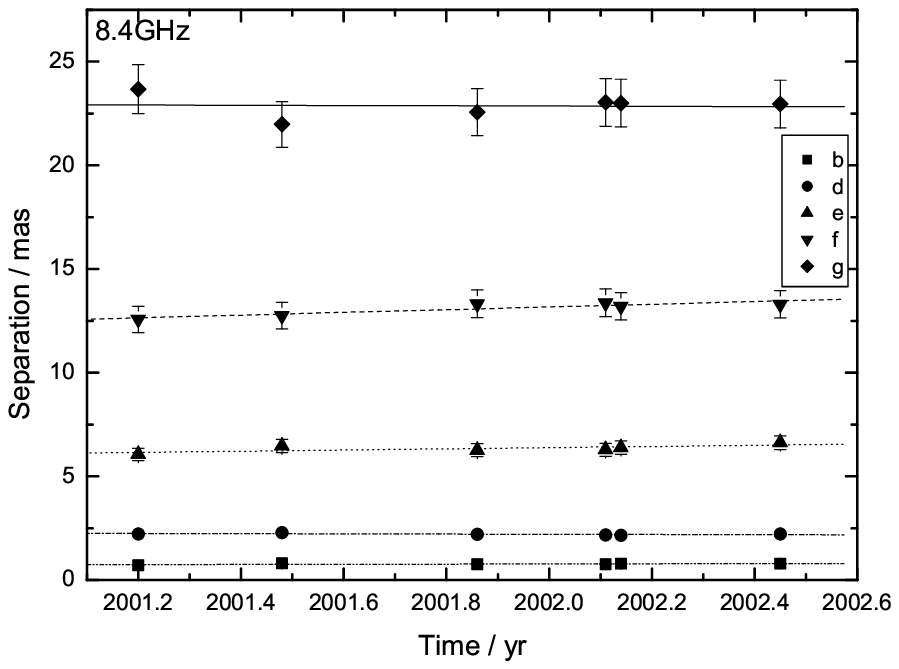}

\caption{The linear fitting to the separations of components b, d, e, f, and g at 2.3 and 8.4 GHz, respectively.
Fitting results are given in Table~\ref{tab:4}.}
   \label{fig:3}
   \end{figure*}

J01, J05, and B05 have studied the jet kinematics in detail, and they showed that the kinematics of 3C~66A is
complicated; some components show superluminal motions, some show the inward (apparently toward the core)
motions, and some show the zero proper motions. We also linearly fit the separations of the jet components at
2.3 and 8.4 GHz (Fig.~\ref{fig:3}), where the fitting results are listed in Table~\ref{tab:4}. Since the
observations at 22.2 GHz were at two short-separated epochs, we do not fit the separations of the jet components
at 22.2 GHz. We can see from Fig.~\ref{fig:3} and Table~\ref{tab:4} that there are no obvious proper motions of
the jet components at 2.3 and 8.4 GHz within the uncertainties. B05 also find no obvious superluminal motions of
the components B05-B2 (we add ``B05'' before the corresponding component name used in their paper, this kind of
nomenclature is also used for the components in J01 and J05), B05-B3, B05-C2, and B05-C3 except that B05-C1 has
the apparent velocity 12.1$\pm$8.0c through their three epochs (2003.78, 2003.83, and 2004.08) observations at
22 and 43 GHz.

\begin{table*}
\begin{minipage}{120mm}
\begin{center}\caption{The proper motions of the jet components in 3C~66A during our six epochs.} \label{tab:4} \vspace{2mm}
\begin{tabular}{cccccccccc}\hline\hline
&b$_{8.4}$ &d$_{2.3}$ &d$_{8.4}$&e$_{2.3}$&e$_{8.4}$&f$_{2.3}$&f$_{8.4}$&g$_{2.3}$&g$_{8.4}$\\\hline $\mu\ ({\rm
mas\,yr}^{-1})$&0.038$\pm$0.029&-0.12$\pm$0.10&-0.048$\pm$0.040&0.14$\pm$0.35&0.28$\pm$0.15&-0.10$\pm$0.57&0.66$\pm$0.16&0.60$\pm$1.25&-0.06$\pm$0.61\\\hline
$\beta_{app}$&1.02$\pm$0.78&-3.21$\pm$2.68&-1.28$\pm$1.07&3.75$\pm$9.37&7.50$\pm$4.02&-2.68$\pm$15.27&17.68$\pm$4.28&16.07$\pm$33.48&-1.61$\pm$16.34\\\hline

\end{tabular}
\end{center}
\end{minipage}
\end{table*}

\begin{figure*}
   \includegraphics[0,0][332,237]{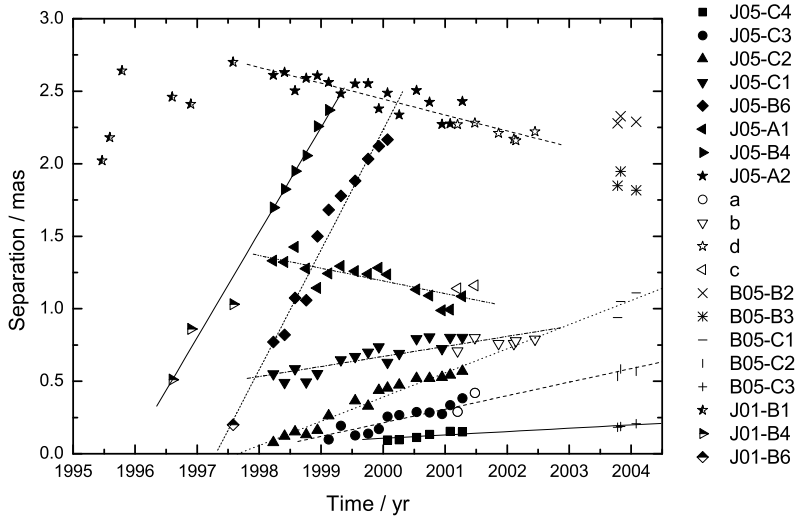}
\caption{Identification of our jet components with the jet components detected in J01, J05, and B05. The open
symbols represent our components (a, b, c, and d), filled symbols represent the components (J05-C4, J05-C3,
J05-C2, J05-C1, J05-B4, J05-B6, J05-A1, and J05-A2) in J05, the half-filled symbols represent the components
(J01-B1, J01-B4, and J01-B6) in J01, and the line-composed symbols represent the components (B05-B2, B05-B3,
B05-C1, B05-C2, and B05-C3) in B05. See text for more details.}
   \label{fig:4}
   \end{figure*}

To show a superluminal motion of 1c in 3C~66A, a proper motion of at least 0.037 mas\,yr$^{-1}$ should be
detected. But the errors of proper motions in Table~\ref{tab:4} are all greater than 0.037 mas\,yr$^{-1}$ except
for component b at 8.4 GHz, which means that the non-detection of superluminal motion may be due to the short
time baseline of our observations, so we studied the kinematics of the jet components by linking our
observations with the observations of J01, J05, and B05. We investigated the relations among our components and
the components in B05, J01, and J05 by extrapolating the detected proper motions in J01 and J05 because of their
more frequent and longer-spanned observing epochs. Figure~\ref{fig:4} shows the relations of these components,
where the data points of components a and c are from our observations at 22.2 GHz, the data points of components
b and d are from our observations at 8.4 GHz, and other data points are from observations at 43 GHz in J01, J05,
and B05. Although there is a difference in the separations from the core of the same component at 8.4/43 GHz and
at 22.2/43 GHz probably due to the SSA, we show in Sect. 3.3 that such a positional difference is small (less
than 10\% of the separation of the jet component relative to the core), so combining the data at 8.4, 22.2, and
43 GHz together to discuss the proper motion barely affects on our results. We can see from Fig.~\ref{fig:4}
that J05-C4, J05-C3, J05-C2, J05-C1, J05-B6, J05-A1, J05-B4, and J05-A2 may be identified with B05-C3, a (and
B05-C2), B05-C1, b, J01-B6, c, J01-B4, and d, respectively. We also try the linear fitting to the separations of
these components based on our identifications and give the results in Table~\ref{tab:5}.

\setlength{\tabcolsep}{3pt}
\begin{table*}
\begin{minipage}{120mm}
\begin{center}\caption{The proper motions by linearly-fitting to the separations of the jet components in Fig.~\ref{fig:4} by assuming the identifications of these jet components. See text for details.} \label{tab:5} \vspace{2mm}
\begin{tabular}{ccccccccc}\hline\hline
&J05-C4/B05-C3 &J05-C3/a/B05-C2 &J05-C2/B05-C1&J05-C1/b&J05-B6/J01-B6&J05-A1/c&J05-B4/J01-B4&J05-A2/d\\\hline
$\mu\ ({\rm
mas\,yr^{-1}})$&0.023$\pm$0.003&0.093$\pm$0.006&0.17$\pm$0.01&0.069$\pm$0.01&0.83$\pm$0.04&-0.088$\pm$0.016&0.72$\pm$0.04&-0.11$\pm$0.01\\\hline
$\beta_{app}$&0.62$\pm$0.08&2.49$\pm$0.16&4.55$\pm$0.27&1.85$\pm$0.27&22.23$\pm$1.07&-2.36$\pm$0.43&19.28$\pm$1.07&-2.95$\pm$0.27\\\hline

\end{tabular}
\end{center}
\end{minipage}
\end{table*}

According to Fig.~\ref{fig:4}, we speculate that J01-B1 may be identified with J05-A2. Although B05-B3 may be
identified with J05-A2 according to the motion trend of J05-A2, B05-B2 is more possibly identified with J05-A2
because their position angles are almost the same, $\sim-173\degr$. It is also possible that J05-A1 or J05-C1
may be identified with B05-C1. Obviously future high-resolution VLBI observations are needed to confirm these
identifications.

\subsection{Frequency-dependent positional difference of the VLBI components}

It can be seen from Table~\ref{tab:2} that the separations of components b and d from the core at 8.4 GHz are
smaller than the ones at 22.2 GHz, and the separations of component d at 2.3 GHz are also smaller than at 8.4
GHz; i.e., there is a difference of position from the core of the same jet component at different frequencies.
There is no obvious frequency-dependent positional difference for the outer jet components, which may be due to
their weak emissions and thus large positional uncertainties. According to Lobanov (\cite{Lobanov98}), the SSA
can cause the offset of the position from the jet apex of the ultracompact VLBI core at different frequencies
due to the frequency-dependent opacity effects since the VLBI core is observed at the location where the optical
depth of SSA is 1. Consequently the separations from the core of the optically-thin jet component at different
frequencies will be different. In the following, we adopt this model to estimate the separations of 3C~66A core
from the jet apex at different frequencies.

Under the assumption that the magnetic field and electron density within the ultracompact VLBI core region
decrease with r, the distance from the conical jet apex, as $B\propto r^{-m}$ and $N\propto r^{-n}$ (K\"{o}nigl
\cite{Konigl81}), the projected angular distance of the core from the jet apex on the sky plane at the observing
frequency $\nu$ can be expressed as (Lobanov \cite{Lobanov98}):
$${\rm r}_{a,proj}\ ({\rm mas})=\triangle
r_{mas}(\nu_1,\nu_2)\frac{\nu_1^{1/k_r}\nu_2^{1/k_r}}{\nu_2^{1/k_r}-\nu_1^{1/k_r}}\cdot\nu^{-1/k_r},\eqno(1)$$
where $k_r=((3-2\alpha)m+2n-2)/(5-2\alpha)$, and $\triangle r_{mas}(\nu_1,\nu_2)$ is the offset of the compact
core position measured from the jet apex at frequencies $\nu_1$ and $\nu_2$ in the units of mas. This is
eventually equivalent to the observed positional difference in the optically-thin jet component at frequencies
$\nu_1$ and $\nu_2$, so the true, frequency-independent angular distance of the jet component from the jet apex
is:
$$R_{a}({\rm mas})={\rm R}+{\rm r}_{a,proj},\eqno(2)$$
where R is the observed jet separation from the optically-thick core (Col. (4) in Table~\ref{tab:2}) at
frequency $\nu$. By using Eqs. (1) and (2) and adopting $k_r$=1 in the case of the equipartition between
particle and magnetic-field energy densities, the relocated separations ${\rm R}_a$ of the jet component d can
be calculated from the observational data at 2.3 and 8.4 GHz. These results found in Table~\ref{tab:6} are
actually the same within their errors for all six epochs. By substituting values of $\triangle r_{mas}$ between
frequencies 2.3 and 8.4 GHz (listed in Table~\ref{tab:6}) into Eq. (1), we can further estimate the separation
of the core from the jet apex (${\rm r}_{a,proj}$) at 22.2 GHz. These are 0.033 and 0.030 mas at two epochs
2001.20 and 2001.48, respectively. Compared with those estimates of ${\rm r}_{a,proj}$ at 2.3 and 8.4 GHz
(Table~\ref{tab:6}), we can infer from Eq. (2) that the positional differences in R are 0.29 and 0.26 mas
between 2.3 and 22.2 GHz at 2001.20 and 2001.48, and 0.054 and 0.049 mas between 8.4 and 22.2 GHz at 2001.20 and
2001.48, which are consistent with the observed differences ($\triangle {\rm}r_{mas}$) at 2.3/22.2 GHz and
8.4/22.2 GHz (see Table~\ref{tab:2}). Similarly, with the average positional difference of component d between
2.3 and 8.4 GHz of 0.27$\pm$0.03 mas, we can obtain the average separation of the core from the jet apex (${\rm
r}_{a,proj}$) at 43 GHz of 0.020 mas. So the positional differences of the jet components from the core between
8.4 and 43 GHz and between 22.2 and 43 GHz are only 0.08 and 0.01 mas, which are less than 10\% of the measured
separation of component d from the core. Therefore, we can ignore such a frequency-dependent jet-position
difference in the proper motion estimation discussed in Sect. 3.2.

We must emphasise that the formula for calculating the positional difference of the jet component is based on
the assumption that the magnetic field and electron density continuously decrease with r; i.e., there is a
smooth-flow jet structure within the ultracompact VLBI core region. If this is not the case, we still can
observe the positional difference of the optically-thin jet component relative to the core because of the
opacity effect, but its dependency on frequency would not follow that in Eq. (1).

\begin{table*}
\begin{minipage}{120mm}
\begin{center}\caption{The angular separations of the core and component d relative to the jet apex}
\label{tab:6} \vspace{2mm}
\begin{tabular}{cccccccc}\hline\hline
 &2001.20 &2001.48 &2001.86&2002.11&2002.14&2002.45&average\\\hline
$\triangle
r_{mas}$&0.23$\pm$0.032&0.21$\pm$0.036&0.21$\pm$0.028&0.36$\pm$0.028&0.37$\pm$0.032&0.24$\pm$0.036&0.27$\pm$0.032\\
r$_{a,proj}^{2.3}$&0.32$\pm$0.044&0.29$\pm$0.050&0.29$\pm$0.038&0.50$\pm$0.038&0.51$\pm$0.044&0.33$\pm$0.050&0.37$\pm$0.044\\
r$_{a,proj}^{8.4}$&0.087$\pm$0.012&0.079$\pm$0.014&0.079$\pm$0.010&0.14$\pm$0.010&0.14$\pm$0.012&0.090$\pm$0.014&0.10$\pm$0.012\\
R$_{a}$&2.31$\pm$0.016&2.36$\pm$0.024&2.29$\pm$0.022&2.30$\pm$0.022&2.30$\pm$0.016&2.31$\pm$0.024&2.31$\pm$0.021\\\hline
\end{tabular}
\\
\end{center}
Notes:$\triangle r_{mas}$ is defined to be the positional offset of the core relative to the jet apex, and is
equivalent to the observed positional difference of component d at 2.3 and 8.4 GHz, r$_{a,proj}^{2.3}$ and
r$_{a,proj}^{8.4}$ are the estimated separation (from Eq. (1)) of the core from the jet apex at 2.3 GHz and 8.4
GHz, respectively, and R$_{a}$ (from Eq. (2)) is the estimated separation of component d from the jet apex,
which is independent of frequency.
\end{minipage}
\end{table*}

\begin{figure*}
\includegraphics[0,0][280,220]{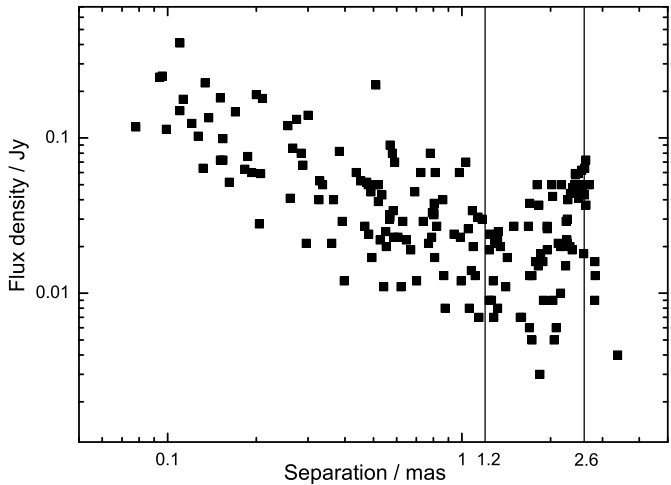}
\caption{The distribution of the flux density of the jet
components against the separation from the core at 43~GHz. The
data are from J05. This figure shows that the flux density of the
jet component starts to increase from about 1.2 mas till about 2.6
mas (indicated by two vertical lines).}

\label{fig:5}
\end{figure*}

\section{Discussion}

Our VLBA maps of 3C~66A show a typical core-jet structure with two bendings at 1.2 and 4~mas from the core.
Furthermore, the VLBA maps at 43 GHz in J05 show that the first bending is a continuous bending beginning at 1.2
mas up to 3 mas from the core. We also find that the flux densities of the VLBI components in J05, B05, and our
data at 22.2 GHz decrease along the jet, but start to increase at the location 1.2 mas up to 2.6 mas from the
core. Such an obvious increase in the flux densities of the VLBI components in J05 beginning from 1.2 mas from
the core can be seen in Fig.~\ref{fig:5}. The coincidence between the place of the first bending and where the
flux densities of the jet components increase suggests that the observed bending may be related to the increase
in the flux densities of the jet. And continuous bending from 1.2 to 3 mas may be due to the continuous decrease
in the viewing angle to the jet components, then the increase of Doppler boosting. It is difficult to judge
whether there is an increase in the flux density to be related to the bending at 4~mas since we cannot find any
obvious increase in the flux density beginning from 4 mas.

Combining our data at 2.3, 8.4, and 22.2 GHz at 2001.48 with the data at 5 GHz at 1995.65 in Taylor et al.
(\cite{Taylor96}), we used the SSA model to fit the spectra of the core for the first time. The best-fit
optically-thin spectral index is -0.08, and B05 give an average spectral index -0.15 from their observations at
22 and 43 GHz at three epochs 2003.78, 2003.83, and 2004.08. The difference in the spectral index may be due to
the different core activities at different observing epochs. By examining the light curves from the UMRAO and
the Mets\"{a}hovi radio observatory (see Ter\"{a}sranta et al. \cite{Terasranta98} and Ter\"{a}sranta et al.
\cite{Terasranta05}), we find that the observing epoch 2001.48 is near the radio burst at 2001.20. The
observations in B05 at three epochs were also close to a radio burst at 2004.0, but the amplitude of this radio
burst is lower than the one at 2001.20. It may be possible that there are more high-energy electrons in the core
region at 2001.48 than at three epochs in B05, resulting in a flatter spectrum at 2001.48 than those at 2003.78,
2003.83, and 2004.08.

We fit the separations of the jet components of 3C~66A at 2.3 and 8.4 GHz for six epochs, and did not detect any
proper motions of these jet components during our observations, which may be due to our short coverage time. We
then studied the relations between our jet components and the components in J01, J05, and B05, and give the
possible identifications of these jet components in Fig.~\ref{fig:4}. We can see from Fig.~\ref{fig:4} that our
jet components are not stationary if our identifications are right. The components J05-A1 and J05-A2 in J05 show
the apparently inward motion to the core, it may be due to the viewing angle to the jet component becoming small
when the jet component moves back toward the line of sight. This is also evidence that the increase in the flux
densities of the jet components from 1.2 to 2.6~mas may be due to decreasing viewing angles to these jet
components.

The average positional difference in component d from the core at 2.3 and 8.4 GHz is 0.27~mas. Lobanov
(\cite{Lobanov98}) give the positional differences of the optically-thin jet components from the core at 2.3 and
8.4 GHz of 3C~395 (z=0.635), 4C~39.25 (z=0.699), and 1038+528A (z=0.678) about 0.60, 0.50, and 0.50 mas,
respectively, which are bigger than the one, 0.27 mas, in 3C~66A. The value of the observed positional
difference of the jet component from the core is determined by the magnetic field and electron density in the
core region, redshift, and viewing angle to the jet. The observed positional difference of the jet component
will be bigger with the stronger magnetic field and higher electron density in the core region, the lower
redshift, and the larger viewing angle. The redshift of 3C~66A is lower than 3C~395, 4C~39.25, and 1038+528A,
but the magnetic field in the core region and the viewing angle are smaller than the ones in 3C~395, 4C~39.25,
and 1038+528A (Lobanov \cite{Lobanov98}), which may explain why the observed positional difference in the
component d in 3C~66A is smaller than the ones in 3C~395, 4C~39.25, and 1038+528A.

\section{Conclusions}

We present results of VLBA observations at 2.3, 8.4, and 22.2 GHz during six epochs (the observations at 22.2
GHz were only performed in the first two epochs). We find from our VLBA maps that there are two bendings at 1.2
and 4 mas from the core. We identify the brightest k component with the core from our simultaneous spectral data
that show an inverse. Combining the existing VLBI data at 5 GHz, we use the SSA model to fit the core spectrum
at epoch 2001.48 for the first time, and obtain the best-fit optically-thin spectral index $-0.08$, the magnetic
filed in the core region 1.1 mG, and the maximum flux density 0.66 Jy at the turnover frequency of 6.52 GHz. We
did not detect any obvious proper motions in jet components during our observing epochs, which may be due to the
short observing time. We give a possible identification of our jet components with the jet components in J01,
J05, and B05 by analysing the proper motions of these jet components. The positional differences in component d
from the core at different frequencies are consistently seen at all the six epochs.

\begin{acknowledgements}
This research has made use of data from the University of Michigan Radio Astronomy Observatory which has been
supported by the University of Michigan and the National Science Foundation. This work was supported in part by
the National Natural Science Foundation of China (grants 10573029, 10625314, and 10633010) and the Knowledge
Innovation Program of the Chinese Academy of Sciences (Grant No. KJCX2-YW-T03), and sponsored by the Programme
of Shanghai Subject Chief Scientist (06XD14024). Z.-Q. Shen acknowledges the support by the One-Hundred-Talent
Programme of the Chinese Academy of Sciences
\end{acknowledgements}

\longtab{2}{
\begin{longtable}{ccccccccc}
\caption{The results of model-fitting at six epochs} \\
 \hline\hline Freq. (GHz)& Comp.& S (Jy)& R (mas)&
$\theta$ (deg)& Major (mas)& Axial Ratio& $\phi$ (deg)& T$_b$ (k)\\
(1)& (2)& (3)& (4)& (5)& (6)& (7)& (8)& (9)\\\hline
\endfirsthead
\caption{Continued.}\\
\hline\hline Freq. (GHz)& Comp.& S (Jy)& R (mas)&
$\theta$ (deg)& Major (mas)& Axial Ratio& $\phi$ (deg)& T$_b$ (k)\\
(1)& (2)& (3)& (4)& (5)& (6)& (7)& (8)& (9)\\\hline
\endhead
\hline
\endfoot
\multicolumn{9}{c}{Epoch 2001.20, \, $\chi^2_{\nu, 2.3}=2.36$,\, $\chi^2_{\nu, 8.4}=1.76$,\, $\chi^2_{\nu,
22}=2.22$}\\\hline 2.3& k& 0.517$\pm$0.052&
0& 0& 0.37$\pm$0.04& 1.0&0 &$1.26\times10^{12}$\\
&d& 0.206$\pm$0.021& 1.99$\pm$0.03& -176.1$\pm0.6$& 0.68$\pm$0.03& 1.0&
0&$1.48\times10^{11}$\\
&e& 0.055$\pm$0.006& 5.85$\pm$0.29& 171.5$\pm$0.04& 3.27$\pm$0.05& 1.0&
0&$1.71\times10^{9}$\\
&f& 0.045$\pm$0.005& 12.69$\pm$0.63& 164.3$\pm$0.6& 3.66$\pm$0.07&
1.0& 0&$1.11\times10^{9}$\\
&g& 0.034$\pm$0.0034& 20.01$\pm$1.00& 166.5$\pm$0.4& 10.03$\pm$0.31&
1.0& 0&$1.12\times10^{8}$\\
8.4& k& 0.564$\pm$0.057& 0& 0& 0.023$\pm$0.010& 1.0& 0&$2.66\times10^{13}$\\
& b& 0.143$\pm$0.014& 0.71$\pm$0.01& -166.0$\pm0.1$& 0.19$\pm$0.01&
1.0& 0&$9.89\times10^{10}$\\
& d& 0.105$\pm$0.011& 2.22$\pm$0.01& -174.7$\pm$0.1& 0.59$\pm$0.01&
1.0&0&$7.53\times10^{9}$\\
&e&0.028$\pm$0.003& 6.06$\pm$0.30& 170.2$\pm$0.6& 3.44$\pm$0.15&
1.0&0&$5.91\times10^{7}$\\
&f& 0.039$\pm$0.004& 12.57$\pm$0.63& 165.3$\pm$0.5& 5.16$\pm$0.25&
1.0& 0&$3.66\times10^{7}$\\
&g& 0.017$\pm$0.0017& 23.68$\pm$1.18& 159.7$\pm$0.8& 7.03$\pm$0.62&
1.0& 0&$8.59\times10^{6}$\\
22.2& k& 0.626$\pm$0.12& 0& 0& 0.025$\pm$0.002& 1.0& 0&$3.58\times10^{12}$\\
&a& 0.160$\pm$0.032& 0.29$\pm$0.01& -165.0$\pm$0.2& 0.14$\pm$0.01&
1.0& 0&$2.92\times10^{12}$\\
&b&0.041$\pm$0.0082& 0.74$\pm$0.02& -168.9$\pm$0.3& 0.042$\pm$0.01&
1.0& 0&$2.31\times10^{10}$\\
&c& 0.054$\pm$0.011& 1.14$\pm$0.01& -165.5$\pm$0.2& 0.33$\pm$0.01& 1.0& 0&$1.77\times10^{9}$\\
&d&0.089$\pm$0.022&2.27$\pm$0.01&-173.4$\pm$0.3&0.53$\pm$0.01&1.0&0&1.13$\times10^{9}$\\\hline
\multicolumn{9}{c}{Epoch 2001.48,\, $\chi^2_{\nu, 2.3}=1.84$,\, $\chi^2_{\nu, 8.4}=1.76$,\, $\chi^2_{\nu,
22}=1.83$}\\\hline 2.3& k&
0.481$\pm$0.048& 0& 0& 0.35$\pm$0.05& 1.0& 0&$1.31\times10^{12}$\\
&d& 0.205$\pm$0.021& 2.07$\pm$0.03& -179.4$\pm$0.3& 0.45$\pm$0.07&
1.0& 0&$3.37\times10^{11}$\\
&e& 0.055$\pm$0.006& 6.85$\pm$0.34& 166.6$\pm$0.3& 3.62$\pm$0.06&
1.0& 0&$1.40\times10^{9}$\\
&f& 0.048$\pm$0.005& 13.95$\pm$0.70& 165.0$\pm$0.3& 5.10$\pm$0.06&
1.0& 0&$6.14\times10^{8}$\\
&g& 0.019$\pm$0.0019& 23.74$\pm$1.19& 165.1$\pm$0.2& 5.01$\pm$0.36&
1.0& 0&$2.52\times10^{8}$\\
8.4&k& 0.637$\pm$0.064& 0& 0& 0.12$\pm$0.01& 1.0& 0&$1.10\times10^{12}$\\
&b& 0.161$\pm$0.016& 0.80$\pm$0.01& -166.7$\pm$0.1& 0.30$\pm$0.02&
1.0& 0&$4.47\times10^{10}$\\
&d& 0.120$\pm$0.012& 2.28$\pm$0.02& -175.2$\pm$0.1& 0.58$\pm$0.02&
1.0& 0&$8.91\times10^{9}$\\
&e&0.038$\pm$0.004&6.458$\pm$0.32&168.6$\pm$0.4&3.33$\pm$0.07&1.0&0&$8.56\times10^{7}$\\
&f& 0.048$\pm$0.005& 12.75$\pm$0.64& 163.3$\pm$0.9& 7.16$\pm$0.29&
1.0& 0&$2.34\times10^{7}$\\
&g& 0.001$\pm$($>$0.001)& 21.98$\pm$1.10& 164.8$\pm$0.1& 2.49$\pm$0.12&
1.0& 0&$4.03\times10^{6}$\\
22.2&k& 0.624$\pm$0.12& 0& 0& 0.097$\pm$0.003& 1.0& 0&$2.37\times10^{11}$\\
&a& 0.143$\pm$0.029& 0.42$\pm$0.01& -165.7$\pm$0.2& 0.21$\pm$0.01&
1.0& 0&$1.16\times10^{10}$\\
& b& 0.002$\pm$($>$0.002)& 0.81$\pm$0.02& -165.0$\pm$0.3& 0.20$\pm$0.01&
1.0& 0&$1.79\times10^{8}$\\
&c&0.052$\pm$0.010& 1.16$\pm$0.01& -168.8$\pm$0.1& 0.35$\pm$0.02& 1.0& 0&$1.52\times10^{9}$\\
&d&0.064$\pm$0.013&2.31$\pm$0.01&-173.7$\pm$0.1&0.45$\pm$0.01&1.0&0&$1.13\times10^{9}$\\\hline
\multicolumn{9}{c}{Epoch 2001.86,\, $\chi^2_{\nu, 2.3}=1.12$,\, $\chi^2_{\nu, 8.4}=1.11$}\\\hline 2.3&k&
0.591$\pm$0.060& 0& 0& 0.53$\pm$0.02& 1.0& 0&$7.00\times10^{11}$\\
&d&0.226$\pm$0.023& 2.0$\pm$0.02& -176.9$\pm$0.2& 0.92$\pm$0.04&
1.0&0&$8.89\times10^{10}$\\
&e&0.051$\pm$0.005& 6.43$\pm$0.32& 169.3$\pm$0.3& 2.93$\pm$0.09&
1.0& 0&$1.98\times10^{9}$\\
&f&0.060$\pm$0.006& 12.72$\pm$0.76& 165.1$\pm$0.2& 5.30$\pm$0.10&
1.0& 0&$7.11\times10^{8}$\\
&g&0.028$\pm$0.0028& 22.01$\pm$1.10& 165.7$\pm$0.4& 8.35$\pm$0.34&
1.0& 0&$1.34\times10^{8}$\\
8.4&k&0.632$\pm$0.063& 0& 0& 0.10$\pm$0.01& 1.0& 0&$1.58\times10^{12}$\\
&b&0.141$\pm$0.014& 0.76$\pm$0.01& -166.4$\pm$0.2& 0.27$\pm$0.01&
1.0& 0&$4.83\times10^{10}$\\
&d& 0.117$\pm$0.012& 2.21$\pm$0.02& -174.0$\pm$0.1& 0.71$\pm$0.01&
1.0& 0&$5.79\times10^{9}$\\
&e& 0.044$\pm$0.004& 6.26$\pm$0.31& 169.6$\pm$0.5& 4.09$\pm$0.21&
1.0& 0&$6.57\times10^{7}$\\
&f& 0.032$\pm$0.003& 13.32$\pm$0.67& 164.8$\pm$0.4& 4.40$\pm$0.21&
1.0& 0&$4.13\times10^{7}$\\
&g& 0.022$\pm$0.0022& 22.57$\pm$1.13& 166.8$\pm$0.9& 8.58$\pm$1.03& 1.0& 0&$7.46\times10^{6}$\\\hline
\multicolumn{9}{c}{Epoch 2002.11,\, $\chi^2_{\nu, 2.3}=1.05$,\, $\chi^2_{\nu, 8.4}=0.98$}\\\hline 2.3&k&
0.522$\pm$0.052& 0& 0& 0.10$\pm$0.08& 1.0& 0&$1.74\times10^{13}$\\
&d&0.248$\pm$0.025& 1.81$\pm$0.02& -174.8$\pm$0.2& 0.94$\pm$0.05&
1.0& 0&$9.35\times10^{10}$\\
&e&0.057$\pm$0.006& 6.27$\pm$0.31& 169.3$\pm$0.2& 3.00$\pm$0.13&
1.0& 0&$2.11\times10^{9}$\\
&f&0.054$\pm$0.005& 13.07$\pm$0.65& 165.0$\pm$0.2& 5.05$\pm$0.12&
1.0& 0&$7.05\times10^{8}$\\
&g& 0.030$\pm$0.0030& 21.68$\pm$1.08& 166.1$\pm$0.4& 8.56$\pm$0.34&
1.0 & 0&$1.36\times10^{8}$\\
8.4&k& 0.555$\pm$0.060& 0& 0& 0.14$\pm$0.01& 1.0&
0&$7.07\times10^{11}$\\
&b&0.148$\pm$0.015& 0.76$\pm$0.01& -166.3$\pm$0.1& 0.32$\pm$0.04&
1.0& 0&$3.61\times10^{10}$\\
&e& 0.118$\pm$0.012& 2.17$\pm$0.02& -173.4$\pm$0.2& 0.70$\pm$0.02&
1.0& 0&$6.01\times10^{9}$\\
&e& 0.042$\pm$0.004& 6.28$\pm$0.31& 170.4$\pm$0.4& 3.89$\pm$0.13&
1.0& 0&$6.93\times10^{7}$\\
&f&0.04$\pm$0.004& 13.37$\pm$0.67& 165.4$\pm$0.4& 5.22$\pm$0.68&
1.0& 0&$3.66\times10^{7}$\\
&g& 0.015$\pm$0.0015& 23.04$\pm$1.15& 167.3$\pm$1.0& 8.24$\pm$0.19& 1.0& 0&$5.52\times10^{6}$\\\hline
\multicolumn{9}{c}{Epoch 2002.14,\, $\chi^2_{\nu, 2.3}=1.17$,\, $\chi^2_{\nu, 8.4}=1.11$}\\\hline
2.3&k&0.402$\pm$0.040& 0& 0& 0.20$\pm$0.06& 1.0& 0&$3.35\times10^{12}$\\
&d&0.243$\pm$0.024& 1.79$\pm$0.03& -176.7$\pm$0.2& 0.92$\pm$0.04&
1.0& 0&$9.56\times10^{10}$\\
&e&0.052$\pm$0.005& 6.19$\pm$0.31& 168.0$\pm$0.3& 2.94$\pm$0.10&
1.0& 0&$2.00\times10^{9}$\\
&f& 0.056$\pm$0.006& 12.83$\pm$0.64& 164.9$\pm$0.3& 5.25$\pm$0.13&
1.0& 0&$6.77\times10^{8}$\\
&g& 0.030$\pm$0.0030& 22.0$\pm$1.1& 166.6$\pm$0.5& 8.35$\pm$0.47&
1.0& 0&$1.43\times10^{8}$\\
8.4&k&0.529$\pm$0.053& 0& 0& 0.14$\pm$0.01& 1.0& 0&$6.74\times10^{11}$\\
&b& 0.137$\pm$0.014& 0.78$\pm$0.01& -166.9$\pm$0.2& 0.33$\pm$0.01&
1.0& 0&$3.14\times10^{10}$\\
&d& 0.118$\pm$0.012& 2.16$\pm$0.01& -173.5$\pm$0.1& 0.56$\pm$0.01&
1.0& 0&$9.39\times10^{9}$\\
&e& 0.043$\pm$0.004& 6.39$\pm$0.32& 170.5$\pm$0.6& 3.72$\pm$0.14&
1.0& 0&$7.76\times10^{7}$\\
&f&0.037$\pm$0.004&13.21$\pm$0.66& 165.3$\pm$0.5& 5.02$\pm$0.29&
1.0& 0&$3.66\times10^{7}$\\
&g& 0.026$\pm$0.0026& 23.0$\pm$1.15 &165.6$\pm$1.5& 10.52$\pm$1.4& 1.0& 0&$5.86\times10^{6}$\\\hline
\multicolumn{9}{c}{Epoch 2002.45,\, $\chi^2_{\nu, 2.3}=1.06$,\, $\chi^2_{\nu, 8.4}=1.05$}\\\hline
2.3&k&0.429$\pm$0.043& 0& 0& 0.30$\pm$0.05& 1.0& 0&$1.59\times10^{12}$\\
&d&0.209$\pm$0.021& 1.98$\pm$0.03& -174.9$\pm$0.2& 0.81$\pm$0.04&
1.0& 0&$1.06\times10^{11}$\\
&e&0.054$\pm$0.005& 6.47$\pm$0.32& 168.6$\pm$0.3& 2.90$\pm$0.08&
1.0& 0&$2.14\times10^{9}$\\
&f&0.056$\pm$0.006& 13.14$\pm$0.66& 165.3$\pm$0.3& 5.32$\pm$0.13&
1.0& 0&$6.59\times10^{8}$\\
&g&0.031$\pm$0.0031& 22.1$\pm$1.10& 165.6$\pm$0.6& 10.12$\pm$0.45&
1.0& 0&$1.01\times10^{8}$\\
8.4&k&0.447$\pm$0.045& 0& 0& 0.20$\pm$0.01& 1.0& 0&$2.79\times10^{11}$\\
&b& 0.136$\pm$0.014& 0.79$\pm$0.02& -165.9$\pm$0.2& 0.37$\pm$0.01&
1.0& 0&$2.48\times10^{10}$\\
&d&0.123$\pm$0.012& 2.22$\pm$0.02& -174.1$\pm$0.2& 0.75$\pm$0.01&
1.0& 0&$5.46\times10^{9}$\\
&e&0.046$\pm$0.005& 6.62$\pm$0.33& 168.4$\pm$0.5& 4.21$\pm$0.14&
1.0& 0&$6.48\times10^{7}$\\
&f&0.045$\pm$0.004& 13.3$\pm$0.66& 164.9$\pm$0.4& 5.27$\pm$0.21&
1.0& 0&$4.04\times10^{7}$\\
&g&0.021$\pm$0.0021& 22.96$\pm$1.15& 168.7$\pm$0.8& 7.07$\pm$1.78& 1.0& 0&$1.05\times10^{7}$\\

\label{tab:2}

\end{longtable}

\vspace{-2mm}Notes:(1) Observing frequency; (2) component name; (3) flux density of component;(4) separation
from the core (component k); (5) P.A. (from noth to east); (6) major axis of the elliptical Gaussian component;
(7) ratio of the minor to major axes; (8) position angle of the major axis of the elliptical Gaussian component
from north to east; (9) brightness temperature according to the formula given in Shen et al. (\cite{Shen97}). }

\end{document}